\documentclass[12pt,epsfig,amsmath,amssymb,pstricks]{article}
\usepackage{graphicx}

\begin{document}

\title{\bf Effective dark energy from decoherence}

\author{{Chris Fields}\\ \\
{\it 243 West Spain Street}\\
{\it Sonoma, CA 95476 USA}\\ \\
{fieldsres@gmail.com}}
\maketitle

\begin{abstract} Within the quantum Darwinist framework introduced by W. H. Zurek ({\em Nat. Phys.}, 5:181-188, 2009), observers obtain pointer-state information about quantum systems by interacting with a local sample of the surrounding environment, e.g. a local sample of the ambient photon field.  Because the environment encodes such pointer state information uniformly and hence redundantly throughout its entire volume, the information is equally available to all observers regardless of their location.  This framework is applied to the observation of stellar center-of-mass positions, which are assumed to be encoded by the ambient photon field in a way that is uniformly accessible to all possible observers.  Assuming Landauer's Principle, constructing such environmental encodings requires $(ln2) kT$ per encoded bit.  For the observed 10$^{24}$ stars and a uniform binary encoding of center-of-mass positions into voxels with a linear dimension of 5 km, the free energy required at the current CMB temperature T = 2.7 K is $\sim$ 2.5 $\cdot$ 10$^{-27}$ kg $\cdot$ m$^{-3}$, strikingly close to the observed value of $\Omega_{\Lambda} \rho_{c}$.  Decreasing the voxel size to $(l_{P})^{3}$ results in a free energy requirement 10$^{117}$ times larger.

\end{abstract} 



\section{Introduction} 

The Planck results \cite{planck:13} together with earlier data \cite{tegmark:04, tegmark:06, frieman:08} clearly establish an observational effect of dark energy in our universe, setting the value $\Omega_{\Lambda} = \mathrm{0.69 \pm 0.01}$ for the fractional contribution of dark energy to the critical density.  The source of this dark energy, however, remains unclear.  Here I show that decoherence requires a uniform dark energy in any universe in which measurements of center-of-mass positions of macroscopic objects yield objective, effectively-classical values for all observers regardless of their location.  

Classical general relativity as well as $\Lambda$-CDM cosmology treat stars and other macroscopic objects as having objective, effectively-classical center-of-mass positions.  Consistent with the ``environment as witness'' formulation of decoherence \cite{zurek:04, zurek:05} and its ``quantum Darwinist'' extension to multiple observers \cite{zurek:06, zurek:09, zurek:10, zurek:12, zurek:14, horodecki:15, brandao:15}, observers obtain information about such center-of-mass positions not by interacting with stars or other macroscopic objects directly, but rather by interacting with local, mutually-independent samples of the environment, in particular, local, mutually-independent samples of the ambient photon field.  For these local interactions with the ambient photon field to be mutually non-perturbing, the local encoding of center-of-mass positions must be effectively classical.  Assuming Landauer's principle \cite{landauer:61, landauer:99}, the free energy required for this local, classical encoding of center-of-mass positions is $(ln2) kT$ per bit.  The current observed value for the dark-energy density $\Omega_{\Lambda} \rho_{c} = 5.7 \cdot 10^{-27}$ kg$\cdot$m$^{-3}$ corresponds to the free energy density required to uniformly encode classical center-of-mass positions for $10^{24}$ stars with $\sim 5$ km spatial resolution in all samples of the ambient photon field throughout the observable universe.  Decreasing the spatial resolution of the encoding to the Planck scale results in a factor of $10^{117}$ increase in the free energy required for encoding, suggesting that the well-known discrepancy between effective field theory calculations of the vacuum energy and the observed dark-energy density (e.g. \cite{rugh:02}) may be largely due to unrealistic assumptions about the effective classicality of information at small scales.

\section{Decoherence, the environment as witness and quantum Darwinism}

Decoherence is the apparent loss of quantum coherence from a system that is exposed to an unobserved and uncharacterized environment \cite{zeh:70, zeh:73, zurek:81, zurek:82, joos-zeh:85, zurek:98, zurek:03, schloss:07}.  Following \cite{joos-zeh:85, schloss:07}, decoherence can be viewed as a scattering process characterized by a scattering constant $L$.  Using \cite{schloss:07}, Eq. 3.67 and converting to SI units, $L$ for an object of radius $a$ and dielectric constant $\varepsilon \gg 1$ exposed to an ambient photon field at an effective temperature $T$ can be approximated as:

\begin{equation}
L \sim 10^{32} \cdot a^{6} \cdot T^{9} ~\mathrm{m^{-2} \cdot s^{-1}}.
\end{equation}

The characteristic time for decoherence at a spatial scale $x$ is then (\cite{schloss:07}, Eq. 3.58):

\begin{equation}
\tau_{x} = L^{-1} x^{-2} ~\mathrm{s}.
\end{equation}

Consider the CMB as the ambient field, so $T = \mathrm{2.7 ~K}$, $T^{9} \sim \mathrm{10^{4}}$ and let $x = \mathrm{1 ~m}$.  For an atom of the Bohr radius, we obtain $L \sim \mathrm{1.5 \cdot 10^{-26} ~m^{-2} \cdot s^{-1}}$ and $\tau \sim \mathrm{7 \cdot 10^{25} ~s}$, considerably more than the age of the universe.  For a star of one solar radius, however, $L \sim \mathrm{10^{89} ~m^{-2} \cdot s^{-1}}$ and $\tau \sim \mathrm{10^{-89} ~s}$, i.e. the CMB efficiently decoheres the center-of-mass positions of stars.

W. H. Zurek and colleagues introduced the ``environment as witness'' formulation of decoherence \cite{zurek:04, zurek:05} in recognition of the fact that observers typically obtain effectively-classical information about states of quantum systems not by interacting with the systems directly but rather by interacting with the surrounding environment, e.g. the ambient photon field.  In this formulation of decoherence, the environment selectively encodes information only about the ``pointer'' states of systems, which are the eigenstates of system - environment interactions.  System - environment interactions are entirely observer-independent; these eigenstates are therefore ``objective'' from the perspective of any observer.  The environmentally-encoded pointer state information is, therefore, also objective from the perspective of any observer.  An observer can obtain the encoded pointer-state information, but no other information about a quantum system, via a local interaction with the environment.  This local observer - environment interaction does not disturb either the system or any other parts of the environment and is therefore effectively classical.  The environment is, in this case, effectively a classical information channel from the system to the observer, one that transmits only pointer-state information.

\centerline{\includegraphics[width=15cm]{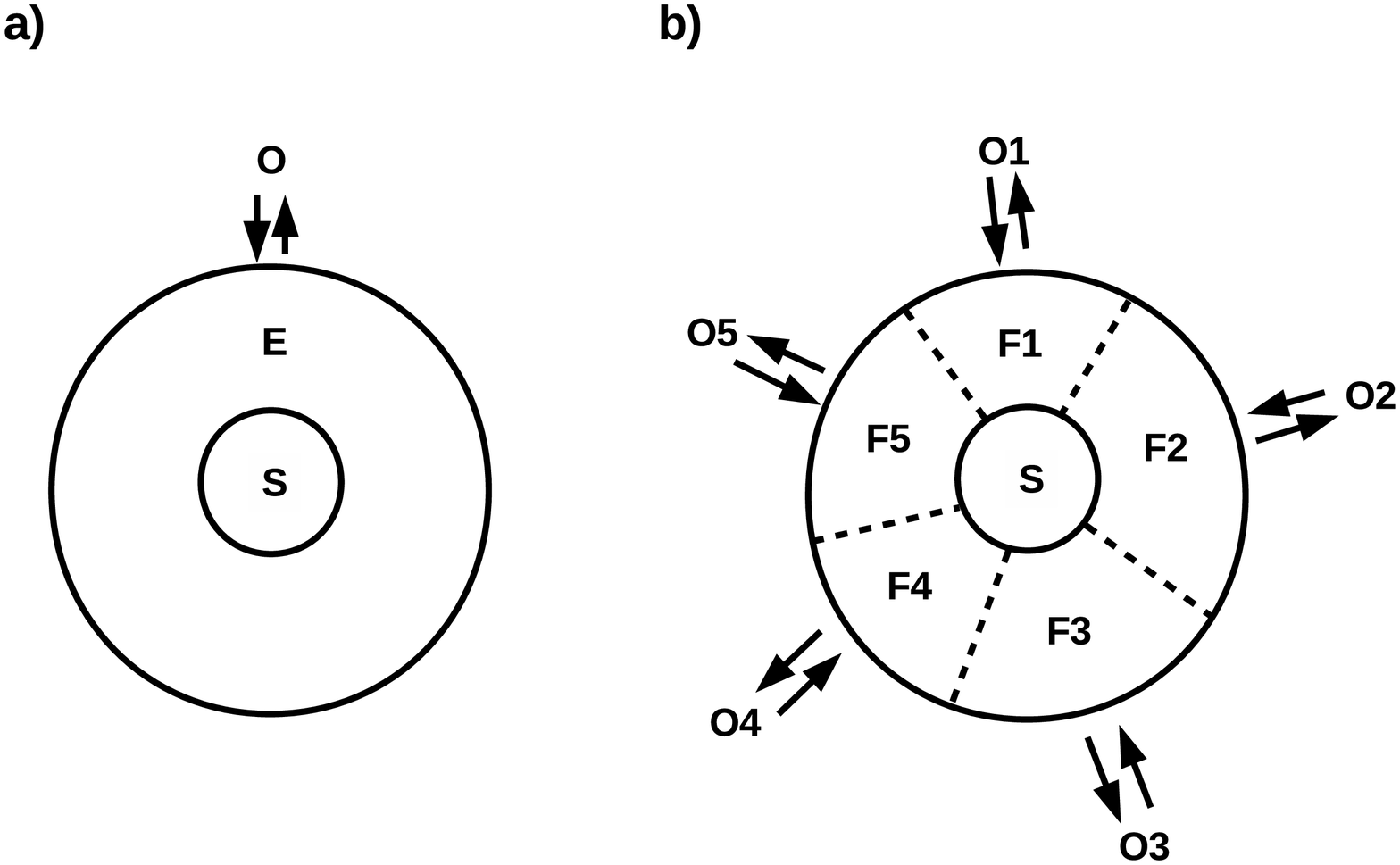}}
\begin{quote}
\textit{Fig. 1}: a) The ``environment as witness'' formulation of decoherence recognizes that an observer $\mathbf{O}$ obtains pointer-state information about a quantum system $\mathbf{S}$ by interacting with the surrounding environment $\mathbf{E}$, e.g. the ambient photon field.  b) Quantum Darwinism extends the environment as witness formulation to multiple observers $\mathbf{O1} \dots \mathbf{ON}$ interacting with local fragments $\mathbf{F1} \dots \mathbf{FN}$ of $\mathbf{E}$, each of which redundantly encodes the same pointer-state information about $\mathbf{S}$.
\end{quote} 

Quantum Darwinism extends this formulation of decoherence to observations made by multiple observers $\mathbf{O1} \dots \mathbf{ON}$, each of which interacts only with their own local fragment $\mathbf{F1} \dots \mathbf{FN}$ of the environment \cite{zurek:06, zurek:09, zurek:10, zurek:12, zurek:14, horodecki:15, brandao:15}.  The environment objectively encodes pointer-state information for any quantum systems embedded in it in a way that is both uniform and massively redundant, making this information equally accessible to the many mutually non-interacting observers, each of whom can obtain the encoded information without disrupting either the systems themselves or the encodings accessed by other observers.  Other than that they each interact only with their own local fragment of the environment, no physical restrictions are placed on the observers in this picture; in particular, they are not restricted in either their location with respect to the observed systems or the resolution with which they interact with the environment.  The resolution with which the environment encodes pointer-state information about any particular system depends only on the system-environment interaction and does not, in particular, depend in any way on any of the observer - environment interactions.

\section{Quantum Darwinism in a cosmological setting}

Decoherence is routinely (e.g. \cite{martineau:06, kiefer:09, hartle:11, calzetta:12, tegmark:12} among others) but by no means universally (e.g. \cite{martin:12, canate:13}) invoked in cosmological settings to explain both unidirectional time evolution and apparently classical values for degrees of freedom such as center-of-mass positions or velocities.  However, the question of how decoherence makes the \textit{same} information about center-of-mass positions, velocities or any other effectively classical degrees of freedom available to multiple, independent observers -- the question of encoding redundancy raised by quantum Darwinism -- has yet to be addressed explicitly.  The fact that human observers currently obtain all information about center-of-mass positions, velocities and other properties of objects of cosmological interest through interactions with the ambient photon field is consistent with both primary assumptions of quantum Darwinism: that pointer-state information is redundantly encoded by states of the same environment that is responsible for decoherence and that observers obtain such information solely by interacting with local fragments of the decohering environment.  We assume, therefore, that the ambient photon field is the environment $\mathbf{E}$ responsible for decoherence throughout the human-observable universe $U_{H}$ from the photon decoupling time if not earlier, and that the ambient photon field encodes the decohered center-of-mass positions, velocities and other observable properties of objects of cosmological interest with sufficient redundancy to allow their detection by multiple observers. 

Two epochs of decoherence by the ambient photon field can be distinguished: an initial epoch in which the matter in $U_{H}$ consists solely of populations of identical particles, and a second, later epoch in which at least some matter is gravitationally bound into distinct, individually-identifiable, effectively-classical macroscopic systems such as stars.  In the former epoch, global states of the universe are symmetric under swaps of one particle for another within a single population, e.g. swaps of one electron for another electron or of one H atom for another H atom.  During this epoch of swap-symmetric global states, the ambient photon field encodes information about the instantaneous spatial distribution of particles within each population, but no information about which particular particle, e.g. each particular electron or H atom, occupies which position.  In the latter epoch, this swap symmetry is broken; indeed ``classicality'' can be defined by the twin requirements of unidirectional time evolution and the presence of distinct, non-swappable ``objects'' that maintain their individual identities through time while changing their states.  During this second epoch, the ambient photon field encodes information not only about the instantaneous spatial distribution of systems, but also about which particular system occupies which position.  This non-swap-symmetric epoch can be taken to have begun at least by the time of earliest star formation, and it continues to the present.  It is only in this epoch that comparisons between records kept by different observers are possible -- observers and recordings are distinct entities and hence are not swap-symmetric -- and hence only here that the encoding redundancy required by quantum Darwinism is relevant.  

Quantum Darwinism is concerned only with pointer-state encodings that are sufficiently redundant that any observer can access the encoded information by interacting with a local fragment of the environment.  While many encodings may be redundant over small scales, e.g. redundant for human observers on Earth, encodings of cosmological interest must be redundant over cosmological scales.  For the present purposes, we regard an encoding of pointer-state information in the ambient photon field of $U_{H}$ as ``sufficiently redundant'' to be of cosmological interest if any observer $X$ located anywhere within $U_{H}$ can access the encoded information by interacting with a local fragment of the ambient photon field of her own observable universe $U_{X}$.  Encodings of center-of-mass positions, in particular, are sufficiently redundant to be of cosmological interest if any observer $X$ located anywhere within $U_{H}$ can access, by interacting locally with the ambient photon field of $U_{X}$, the encodings of those positions that are within the overlap zone $U_{X} \cap U_{H}$ as shown in Fig. 2.  Systems with center-of-mass position encodings that are sufficiently redundant in this sense are considered to be systems or ``objects'' of cosmological interest.  Informally, this requirement of a sufficiently redundant center-of-mass position encoding corresponds to the requirement that systems of cosmological interest have ``ontic'' or ``objective'' center-of-mass positions that are independent of the locations of or spatial coordinate systems used by observers anywhere within our observable universe.

\centerline{\includegraphics[width=15cm]{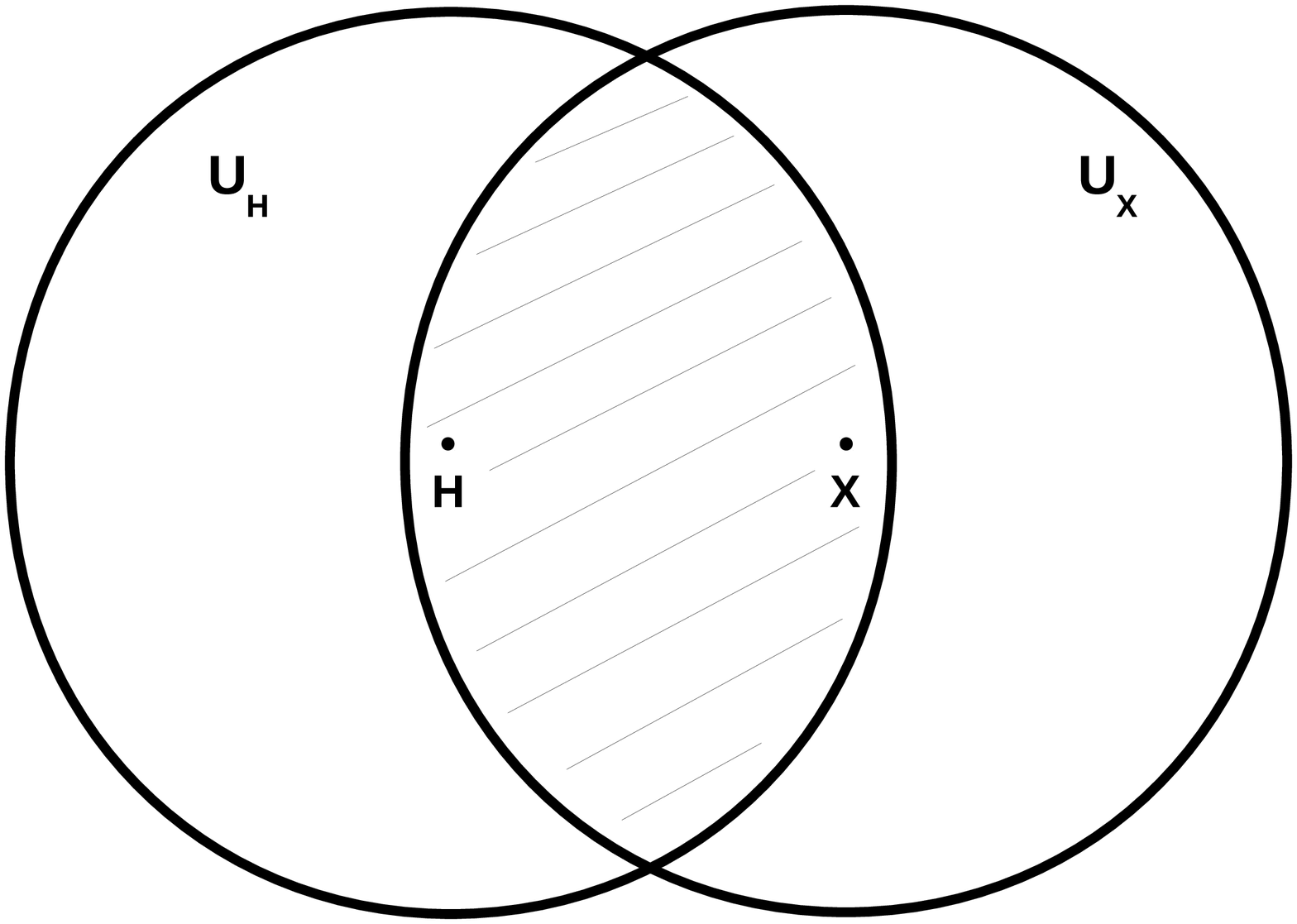}}
\begin{quote}
\textit{Fig. 2}:  An encoding is considered sufficiently redundant if it is accessible to any observer $X$ located anywhere within the human-observable universe $U_{H}$.  A center-of-mass position encoding is sufficiently redundant if any observer $X$ can access, by interacting locally with the ambient photon field of $U_{X}$, the encodings of those positions that are within the overlap zone $U_{X} \cap U_{H}$ (hatched region).
\end{quote} 

Let $N_{O}$ be the number of objects of cosmological interest, as defined by sufficient encoding redundancy, observable within $U_{H}$.  The center-of-mass positions of these $N_{O}$ systems must be encoded within the ambient photon field $\mathbf{E}$ in a way that explicitly specifies both the location of the center of mass of each system, at some spatial resolution $l_{V}$, and the fact that each center of mass is located nowhere else, i.e. that each center of mass has a unique location.  The minimal encoding that meets these two requirements divides $U_{H}$ into non-overlapping three-dimensional voxels of uniform volume $l_{V}^{3}$ and writes $N_{O}$ bits within each voxel, with the $k^{th}$ bit set to `1' within a voxel if and only if the center of mass of the $k^{th}$ object is within that voxel and set to `0' otherwise.  This voxel-based encoding is effectively a database of $N_{O} \cdot N_{V}$ binary records, where $N_{V}$ is the number of voxels.  It is independent of coordinates, and hence may be accessed in the same way by any observer regardless of their location.  Any smaller encoding, e.g. one that provides only a single voxel location for each of the $N_{O}$ systems, fails to meet the requirement of explicitly specifying that no object has multiple center-of-mass positions; however, additional information about each object can be encoded only in the voxel in which the object is located and hence contributes negligibly to bit count of the total encoding.  This explicit, voxel-based encoding is, moreover, the encoding that the ambient photon field presents to human observers: when we look at the sky, we see not only stars, but also the empty space -- and hence the empty voxels -- between them.

\section{Free energy required for encoding}

The environment encodings of pointer-state information described by quantum Darwinism are encodings of \textit{classical} information; it is only by encoding classical information that no-cloning restrictions can be evaded and redundancy achieved \cite{zurek:09}.  Independent accessibility by multiple observers requires, moreover, that these encodings be thermodynamically irreversible.  Landauer's principle therefore applies, requiring the expenditure of $E_{b} = (ln2) kT$ of free energy per encoded bit \cite{landauer:61, landauer:99}.  We can, therefore, associate a free-energy requirement with any environmental encoding of pointer-state information by decoherence.  In particular, we can calculate the free energy required for the explicit, voxel-based encoding of center-of-mass positions of objects of cosmological interest described above.

The number of bits required to unambiguously specify the voxel containing the center-of-mass position of any single object equals the number $N_{V}$ of voxels.  If the positions of $N_{O}$ objects must be simultaneously encoded for each voxel, the total number of bits required is:
\begin{equation}
N_{b} = N_{O} \cdot N_{V}
\end{equation}
as noted above.  Assuming optimal encoding efficiency, the total free energy required to encode these $N_{b}$ bits is:
\begin{equation}
E_{T} = N_{b} \cdot E_{b}, ~ E_{b} = (ln2) kT.
\end{equation}
As $T$ can be considered to be the average temperature of the decohering environment $\mathbf{E}$, i.e. the ambient photon field, it can be equated to the CMB temperature; hence at the present $T = 2.7$ K.  Assuming a uniform encoding throughout the observable universe $U_{H}$, the required free energy density at the present is, in mass units,
\begin{equation}
\rho = (N_{O} \cdot E_{b}) / (l_{V}^{3} \cdot c^{2}),
\end{equation}
where as before $l_{V}$ is the linear dimension of the voxel.  This relation is a power law, as shown in Fig. 3.

\centerline{\includegraphics[width=11cm]{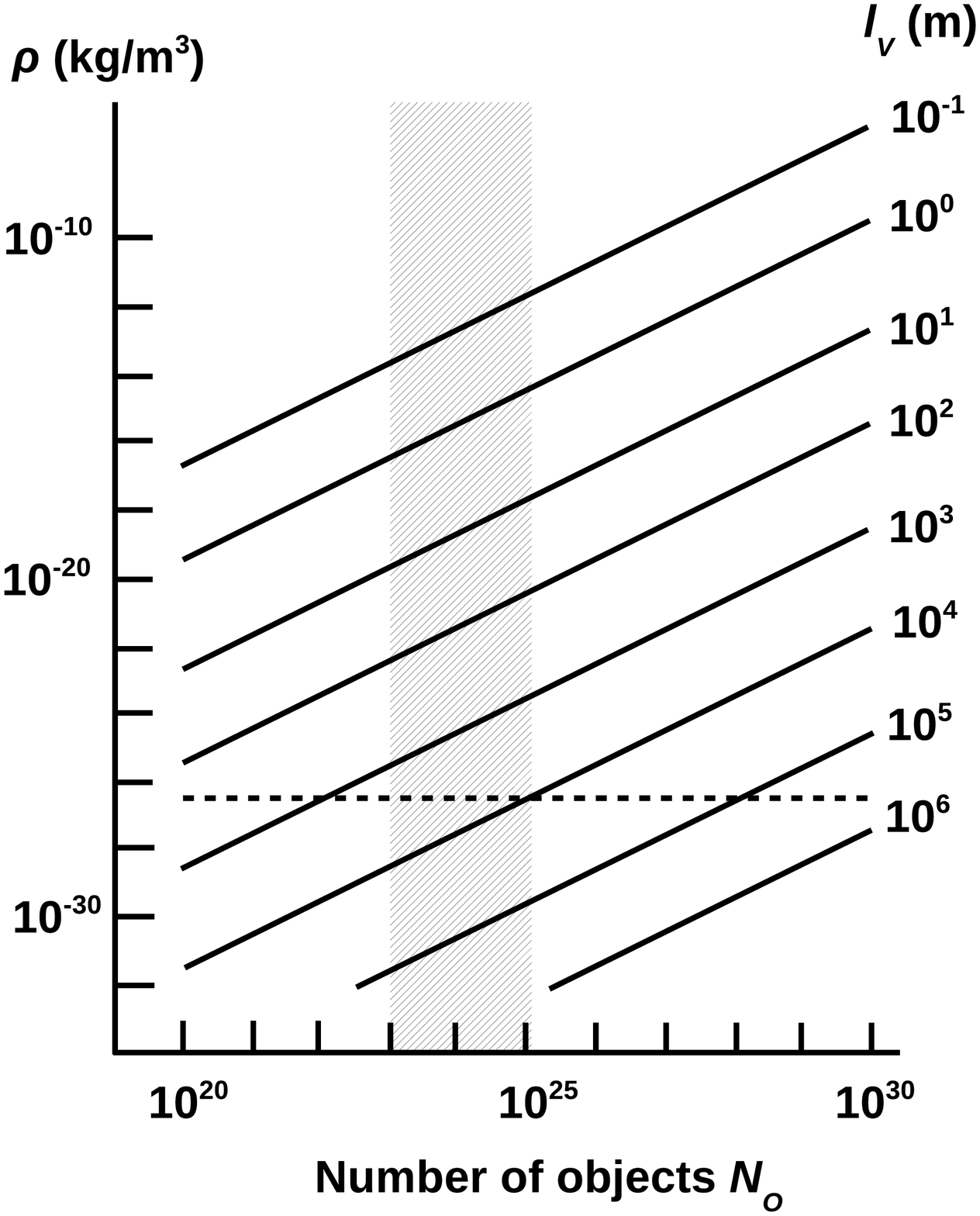}}
\begin{quote}
\textit{Fig. 3}: Plot of the free energy density $\rho$ required for encoding versus the number of binary encoded object positions $N_{O}$ for a range of macroscopic voxel dimensions $l_{V}$.  The shaded box indicates theories in which currently-observed stars have objectively-encoded positions.  The observed value of $\Omega_{\Lambda} \rho_{c} = 5.7 \cdot 10^{-27}$ kg$\cdot$m$^{-3}$ \cite{planck:13} is shown by the dashed horizontal line.
\end{quote}

Observations indicating a ``bottom-heavy'' initial mass function \cite{vanDokkum:10, conroy:12, tortora:13} suggest a total number of $\sim 10^{24}$ stars in our observable universe.  Theories in which the observed stars have objective, classical positions due to environmental decoherence fall, therefore, within the shaded box in Fig. 3.

\section{Physical interpretation of the free-energy density $\mathbf{\rho}$}

In a purely-quantum, decoherence-free universe without a non-unitary physical ``collapse'' process, macroscopic objects would not have observer-independent, effectively-classical center-of-mass positions.  Indeed as pointed out from both physical \cite{zanardi:01, zanardi:04, dugic:06, dugic:08, torre:10, harshman:11, dugic:12} and philosophical \cite{fields:14, kastner:14} perspectives, no quantum-theoretic principle requires that any particular state-space factorization of such a universe is ``preferred'' in any way; hence the notion of an observer-independent ``macroscopic object'' is ill-defined in a purely-quantum, decoherence-free universe without collapse.  In such a universe, there is no objective encoding of classical information and hence no thermodynamic requirement for free energy to support such an encoding.  A purely-quantum, decoherence-free universe without collapse would, therefore, be consistent with $\rho = 0$.  As a positive $\rho$ is purely a product of decoherence, one would not, in particular, expect the zero-point energy of a purely-quantum, decoherence-free vacuum to relate in any way to $\rho$.

As $\rho$ is by definition an \textit{objective} free-energy density required to support an objective encoding of effectively classical center-of-mass positions, however, it must be supplied by some physical source.  If $\rho = 0$ in the presence of only quantum sources, a positive $\rho$ must arise from a classical source.  The only available classical source for such an all-pervasive free-energy density is the classical gravitational field.  \textit{Decoherence alone}, therefore, requires an all-pervasive classical gravitational potential energy density, one equal to the free-energy density $\rho$ required to encode pointer-state information, in particular information specifying effectively-classical center-of-mass positions, uniformly throughout the environment.  The classical cosmological constant is exactly such a density; hence it seems natural to identify $\Omega_{\Lambda} \rho_{c}$ with the energy density $\rho$ required to objectively encode effectively classical center-of-mass position information.  

As shown in Fig. 3, the observed value $\Omega_{\Lambda} \rho_{c} = 5.7 \cdot 10^{-27}$ kg$\cdot$m$^{-3}$ \cite{planck:13} intersects the allowed region for encoding center-of-mass positions for the currently-observed stars at a voxel dimension $l_{V}$ between 1 and 10 km.  Decreasing the voxel dimension to $l_{P}$ results in a value for $\rho$ that is factor of $10^{117}$ times larger than the observed $\Omega_{\Lambda} \rho_{c}$, suggesting that the well-known discrepancy between effective field-theory estimates of the zero-point energy and $\Omega_{\Lambda} \rho_{c}$ may be an artifact of an implausible assumption that classical information remains well-defined and therefore can be encoded at the Planck scale.

\section{Conclusion}

The free-energy cost of encoding classical information has heretofore largely been ignored in discussions of either decoherence or cosmology.  In a decoherence-free universe without collapse, there is no classical information and hence no free-energy cost of encoding.  Such a free-energy cost is, however, inevitably imposed on a quantum universe by decoherence.  As shown here, the free-energy cost to encode effectively-classical center-of-mass positions for the observed $\sim 10^{24}$ stars at a spatial resolution of $\sim 5$ km uniformly throughout the observable universe corresponds to an energy density equal to the observed dark energy density.  The observed dark energy may, therefore, reflect the thermodynamic cost of decoherence and hence the cost of effective classicality.

\section*{Acknowledgement} This work was supported in part by The Federico and Elvia Faggin Foundation.  Thanks to Durmus Ali Demir and Don Hoffman for comments.

\end{document}